\documentclass[floatfix,aps,prapplied,reprint,superscriptaddress,showpacs,showkeys]{revtex4-1}

\usepackage[multidot]{grffile}
\usepackage{amsmath,amssymb}  
\usepackage{color}
\usepackage{graphicx}

\usepackage{titlesec}
\titleformat*{\section}{\normalsize\bfseries\filcenter}
\titleformat*{\subsection}{\normalsize\bfseries\filcenter}
\titleformat*{\subsubsection}{\normalsize\itshape\filcenter}

\draft 

\begin{document}
	
	
	\title{Feedback induced locking in semiconductor lasers with strong amplitude-phase coupling} 

	\author{Jan Hausen}
	\email[]{hausen@campus.tu-berlin.de}
	\affiliation{ Institute of Theoretical Physics, Technische Universit{\"a}t Berlin, Hardenbergstraße 36, 10623 Berlin, Germany}
	
	\author{Bastian Herzog}
	\affiliation{ Institute of Optics and Atomic Physics, Technische Universit{\"a}t Berlin, Hardenbergstraße 36, 10623 Berlin, Germany}
	
	\author{Alexander Nelde}
	\affiliation{ Institute of Optics and Atomic Physics, Technische Universit{\"a}t Berlin, Hardenbergstraße 36, 10623 Berlin, Germany}
	
	\author{Stefan Meinecke}
	\affiliation{ Institute of Theoretical Physics, Technische Universit{\"a}t Berlin, Hardenbergstraße 36, 10623 Berlin, Germany}

	\author{Nina Owschimikow}
	\affiliation{ Institute of Optics and Atomic Physics, Technische Universit{\"a}t Berlin, Hardenbergstraße 36, 10623 Berlin, Germany}

	\author{Kathy L{\"u}dge}
	\email[]{kathy.luedge@tu-berlin.de}
	\affiliation{ Institute of Theoretical Physics, Technische Universit{\"a}t Berlin, Hardenbergstraße 36, 10623 Berlin, Germany}

	\date{\today}
	
	\begin{abstract}
		The influence of optical feedback on semiconductor lasers has been a widely studied field of research due to fundamental interests as well as the optimization of optical data transmission and computing. Recent publications have shown that it is possible to induce a periodic pulsed like output in quantum-dot and quantum-well laser diodes utilizing the locking of the external cavity modes and the relaxation oscillation frequency. We present an in-depth analysis of this effect. We choose submonolayer quantum dots as a gain system, as these provide a relatively strong amplitude-phase coupling, which has proven to be very beneficial for these locking effects to occur. Introducing a new theoretical model we can correctly reproduce the essential features of the gain system and validate them by comparison to our experimental results. From this starting point we can further explore how the staircase behaviour of the oscillation frequency with increasing pump current can be influenced by changing various laser parameters. The staircase behavior is induced by a reordering of the Hopf bifurcations giving birth to the regular pulsed-like oscillations.
	\end{abstract}
	
	\maketitle 
	
	
	\section{Introduction}
	Semiconductor lasers have become a central component of today's optical data transfer due to their well established production and high transmission rates \cite{GRU02,OHT13}. A scheme to severely influence the dynamics of these lasers is direct optical feedback \cite{TKA86}. This method, on the one hand, can be used to stabilize the laser output \cite{DAH87,AHL06,DAH08b}, reduce the linewidth of the laser \cite{DUA18,BRU17,AGR84} or the timing-jitter of a pulsed output \cite{ARS13,OTT12a,JAU16a,LIN11f}. On the other hand, it is also a well established approach to drive the laser into chaotic regimes \cite{CHO84,ALB11,ALN09,OHT99,BOU19}, which can be applied e.g. for the encoding of data \cite{MIR96,HEI02a} or has to be accounted for as a detrimental effect in communication lines. Moreover, lasers subject to feedback are also an interesting candidate to implement computation schemes such as reservoir computing \cite{APP11,BUE17,LAR17,KOE20a}. \\
	Recent works on quantum dot and quantum well lasers under optical feedback have shown that a locking between the external cavity modes and the relaxation oscillation frequency of the continuous wave (CW) solution can lead to a periodic pulsed-like laser output~\cite{TYK16,KEL17b}. Similar effects have also been observed for laser diodes with a 3D gain medium \cite{MOR92,HEL90b,RIT93a,PIE01,ERZ06,KIM15} as well as quantum dash lasers under optoelectronic feedback \cite{KOV19,ISL20}. As the generation mechanism of the periodic states lies within the locking of the internal and external frequency components and therefore requires them to be close together, these oscillations are different from the high frequency oscillations occurring for short optical feedback \cite{TAG94}, pure frequency oscillations \cite{FIS04,ERZ06} or noise driven low frequency oscillations \cite{HOH95,HEI98c}. Moreover, the analysis of the locking effect suggested that a high damping and a high amplitude-phase coupling are very beneficial for the transition from CW lasing to periodic dynamics at experimentally observable feedback-strengths \cite{KEL17b,HAE02}. \\
	In order to further investigate the generality of this feedback effect in terms of the choice of the gain medium, we utilize a laser diode based on submonolayer quantum dots (SMLQDs) \cite{EGO94a,KRE99}. These 0D localization centers have higher areal density than conventional quantum dots \cite{GRU02,CHO13a,JAH12}, and therefore provide a higher gain per unit length \cite{OWS20}.
	Furthermore, these structures exhibit a strong amplitude-phase coupling, i.e. high $\alpha$ factor and preserve advantages such as suppressed carrier diffusion, high gain and phase recovery and low degradation rates \cite{ARS12,LIN16,HER15b,HAR16a,HER16}. Lately, SMLQDs have been successfully implemented as a gain medium in laser diodes as well as external cavity lasers \cite{KRE01,XU04a,DUD17,ALF18}. \\
	In this manuscript we theoretically and experimentally investigate a SMLQD laser diode subject to optical feedback at intermediate feedback lengths ($T = 3.8$\,ns). As expected we observe delay-induced higher-order locking effects also found for quantum wells and quantum dot gain media \cite{TYK16,KEL17b}. This suggests that the feedback is the driving force of the locking as it has a much greater influence than the choice of the gain medium. We support the experimental evidence by developing a relatively simple theoretical model which retains the most relevant time-scales for the locking effects, i.e. the relaxation oscillations \cite{LIN16} and validate it by comparison to experimental findings. Hence, we are able to extend the research on the emergence of the feedback induced pulse-like oscillations \cite{TYK16,KEL17b} by a bifurcation analysis unravelling the underlying generation mechanism of the staircase dependence of the oscillation period on the pump current. Additionally, we discuss how the relaxation oscillations can be altered to influence the staircase and furthermore discuss the role of the amplitude-phase coupling for the occurrence of the pulsed-like states.  
	\section{Experimental Setup and laser device}
	As their name suggests, submonolayer quantum dots (SMLQDs) are grown by repeatedly depositing layers of a lower-band gap material in between a higher band-gap material (e.g. InAs layers on GaAs bulk) \cite{EGO94a,KRE99}. These layers have a thickness below one monolayer (ML). Due to a vertically correlated growth caused by local strain effects, 0D localization
	centers form with a high areal density providing a high gain per unit length \cite{ARS12,LIN16,HER15b}. The investigated SMQLD PIN laser diode was grown utilizing metal organic vapor phase epitaxy. The active section consisted of fifteenfold deposition of 0.73\,ML InAs and 1.13\,ML GaAs respectively. This results in an active section width of 8\,nm, embedded in 10\,nm spacers of GaAs bulk material. To enable single-mode operation, a shallow etched waveguide provided single-mode operation via gain guiding. \\
	\begin{figure}[b]
		\includegraphics{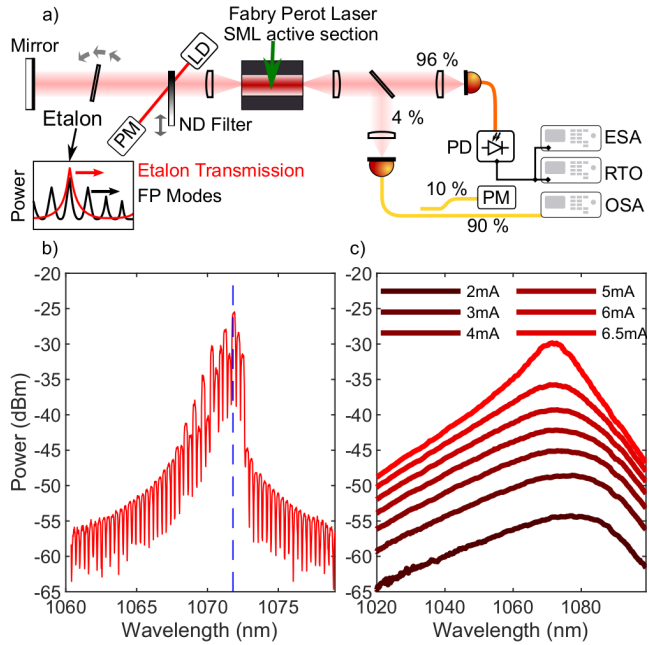}
		\caption{a) Experimental setup utilized to investigate the effect of feedback on a SMLQD laser. An etalon in the feedback-arm (left) enables single-mode operation. The detection-arm (right) includes an electrical spectrum analyser(ESA), an optical spectrum analyser (OSA) and an real-time oscilloscope (RTO). The attenuation of the neutral density (ND) filter is determined using a photo-detector (PM) and a laser diode (LD) b) Optical spectrum (resolution $/0.08$\,nm) of the free-running laser, with the investigated mode to which the etalon was tuned marked by a vertical blue line. c) Coarsely resolved spectrum (resolution mW$/2$\,nm) at different pump currents, indicating the emergence of a lasing mode closely above the threshold of $J_{th} = 6.5$\,mA.  } 
		\label{Fig1}
	\end{figure}
	The optical power spectra were measured with a fiber-coupled optical spectrum analyzer (Hewlett Packard 70952B) and the radio frequency (RF) spectra were obtained using a fiber-coupled high-speed photo-detector (New Focus 1544-A-50) read out by an electrical spectrum analyzer (Rhode \& Schwarz FSC3 Spectrum Analyzer). 
	The Fabry-Perot laser was installed into the external cavity setup schematically shown in Fig.\ref{Fig1}a). By adding an etalon with a transmission linewidth of 0.4\,nm at full-width half maximum into the feedback arm (left), it was possible to select a single lasing mode for the feedback. The optical spectral properties of the free-running device are displayed in Fig.\ref{Fig1}b)-c). A highly resolved spectrum is shown in b), with a blue line indicating the mode for which the maximum feedback was possible. The low resolution spectra for different pump currents in c) clearly indicate the emergence of lasing modes of the free-running device above the threshold of $J_{th} = 6.4$\,mA. 
	\section{Theoretical Model}
\begin{figure}[t]
		\includegraphics{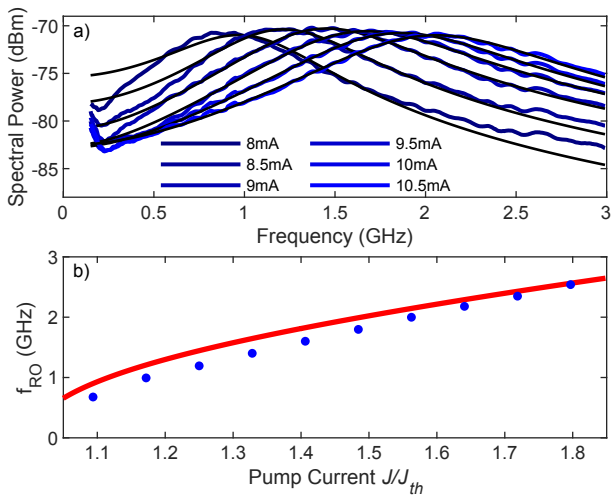}
		\caption{a) Relative intensity noise spectra of the free running laser measured at different pump currents given in the legend, with an noise-equivalent power of $27$\,pW$/\sqrt{\mathrm{Hz}}$. The RIN spectra were fitted according to the routine introduced in \cite{WES89} to extract the relaxation oscillation frequency at each pump current. b) The experimentally determined relaxation oscillations are indicated by blue dots, the red line shows the evolution of the relaxation oscillation frequency with increasing pump current $J$ (normalized to the lasing threshold $J_{th}$) obtained by implementing the model (\ref{Eq:E})-(\ref{Eq:rho}) into DDE-biftool. The Parameters used are shown in Table \ref{Tab1}.} 
		\label{Fig2}
	\end{figure}

	In order to reproduce the dynamics of the investigated device we develop a model based upon the phenomenological approach carried out in \cite{HER16,LIN16}, where the SMLQDs have been divided into subgroups based on their confinement energy and their respective density of states. To arrive at the model presented here, an average over the SML-subgroups is performed, which is justified by the fact that the investigated locking effects occur on a much larger time-scale than the diffusive coupling between the subgroups \cite{LIN16}. The amplitude-phase coupling, including the role of the inactive states, is approximated by a constant large $\alpha$-factor. This also results from the lateral coupling and the size of the inactive reservoir leading to a higher phase-response compared to conventional quantum dots \cite{HER16,OWS20}. Another critical aspect to take into account are the differing relaxation rates of bulk and SMLQD section in the gain. The model describes the dynamics of the device in terms of the complex intra-cavity, dimensionless electric field $E(t)$ within the slowly-varying envelope approximation, the submonolayer quantum dot occupation probability $\rho(t)$ and the bulk reservoir
	charge-carrier density $n(t)$. The resulting equations read:
	\begin{align}
	\label{Eq:E}
	\frac{d}{dt} E(t) &= \frac{1}{2} \left[ g \left(2 \rho(t) - 1\right) (1-i\alpha) -T_{\mathrm{ph}}^{-1} \right] E(t) \nonumber \\
	&+ \frac{K}{2 T_{\mathrm{ph}} } e^{iC} E(t-\tau) + F(t),\\
	\label{Eq:rho}
	\frac{d}{dt} \rho(t) &= -\frac{1}{T_{\rho}}\rho(t) + R\left[ \rho_{\mathrm{eq}}(t) - \rho(t) \right] \nonumber \\ 
	& - g(2\rho(t) - 1)|E(t)|^2, \\
	\label{Eq:n}
	\frac{d}{dt} n(t) &= - \frac{1}{T_{n}}n(t) + \frac{J}{h_{\mathrm{bulk}}} - \frac{2 n_{\mathrm{SML}}}{h_{\mathrm{bulk}}} R\left[ \rho_{\mathrm{eq}}(t) - \rho(t) \right],
	\end{align}
	where $g$ is the differential gain coefficient, $\alpha$ the amplitude-phase coupling coefficient, $T_{\mathrm{ph}}$ the photon lifetime, $K$ the relative feedback strength, $C$ the relative feedback phase, $\tau$ the external cavity roundtrip time (feedback time), $F(t)$ a delta-correlated complex Gaussian Langevin noise term, $T_{\rho}$ the SMLQD lifetime, $h_{\mathrm{bulk}}$ the effective bulk reservoir thickness, $R$ the SMLQD-bulk coupling rate, $\rho_{\mathrm{eq}}$ the SMLQD quasi-equilibrium occupation probability, $J$ the pump current density, $T_n$ the bulk carrier lifetime and $n_{\mathrm{SML}}$ the SMLQD area density. Details on the determination of the SMLQD equilibrium occupation probability $\rho_{eq}$ are given in appendix \ref{AppRho}.
	The SMLQD-bulk coupling rate is modeled with a quadratic dependency on the pump current density:
	\begin{align}
	R = R_0 \left( \frac{J}{J_{\mathrm{tr}}} \right)^2
	\end{align}
	where $J_{\mathrm{tr}}$ is the transparency pump current and $R_0$ the coupling rate at transparency. 
	The feedback term is included as proposed by Lang and Kobayashi \cite{LAN80b,YAN10,ROT07,ORT06,ERN95a,ALS96} and is valid for the small feedback strength ($K < 0.1$) investigated in this manuscript. \\
	In order to adjust the parameters relating the simple SMLQD model eq.(\ref{Eq:E})-(\ref{Eq:n}) and the experimentally investigated device, we measure the relative intensity noise (RIN) spectrum at different pump currents without feedback (multi-mode operation) as shown in Fig.\ref{Fig2}a). Utilizing the fit proposed in \cite{WES89} and indicated by black lines in Fig.\ref{Fig2}a), we are able to extract the relaxation oscillation (RO) frequency of the free-running laser from each RIN spectrum. The resulting experimentally determined relationship of the RO frequency and the pump current is shown by the blue dots in Fig.\ref{Fig2}b).  
	\begin{table}[b]
		\begin{tabular}{cccc}
			\hline
			\hline  
			Symbol & Value & Symbol & Value \\ 
			\hline 
			$J_{\mathrm{th}}$ & $379\cdot 10^{11}$\,ns$^{-1}$\,cm$^{-2}$ & $J_{\mathrm{tr}}$ & $223\cdot 10^{11}$\,ns$^{-1}$\,cm$^{-2}$\\ 
			$\bar{\varepsilon}_{\mathrm{SML}}$  & 0.06\,eV &  $h_{\mathrm{bulk}}$ & $53.0$\,nm \\ 
			$\alpha$ & 5 & $T$ & 300\,K  \\ 
			$n_{\mathrm{SML}}$ & 3.3$\cdot 10^{11}$\,cm$^{-2}$ & $m^{*}$ & $0.07m_e$  \\ 
			$\tau$ & 3.8\,ns & $g$ & 185 \,ns$^{-1}$ \\ 
			$T_{\mathrm{ph}}$ & 9\,ps & $T_{\rho}$ & 0.066\,ns  \\ 
			$T_{\mathrm{n}}$ & 0.5\,ns & $R_{0}$ & 240\,ns$^{-1}$  \\ 
			$K$ & $0$ & $C$ & $0$  \\ 
			\hline 
		\end{tabular} 
		\caption{Parameter values used for the theoretical investigation of the SMLQD device, if not indicated otherwise.}
		\label{Tab1}
	\end{table}
	Implementing the system eq.(\ref{Eq:E})-(\ref{Eq:n}) into the path continuation software DDE-biftool \cite{ENG02} makes it possible to continue the continuous wave (CW) solution in $J$. We execute this in the regime of zero feedback $K = 0$ and identify the RO frequency at each pump current  as the imaginary part of the largest non-trivial (non-zero) Lyapunov exponent. The Lyapunov exponents can be found by applying a linear stability analysis for each point of the CW solution branch. In order to adapt the parameters used in the theoretical model to the investigated device we adjust the parameters so that the simulated relaxation oscillations fit well to the experimentally obtained values. We only slightly adjust the relaxation times and the coupling $R_0$ obtained for a similar laser device presented in Ref. \cite{LIN16}. The parameters chosen in the further course of this manuscript are given in Table \ref{Tab1}, if not indicated otherwise. We achieve a good qualitative agreement between the relaxation oscillation frequency determined via experimental (blue dots) and theoretical results (red line) as displayed in Fig.\ref{Fig2}b). We choose a high $\alpha$ factor to take into account the contribution of the inactive SMLQDs, which is also justified by experimental findings predicting an $\alpha$-factor between 2.5 and 8 for SMLQDs \cite{HER16}. 
	
	\section{Emergence of periodic solutions}
\begin{figure}
		\includegraphics{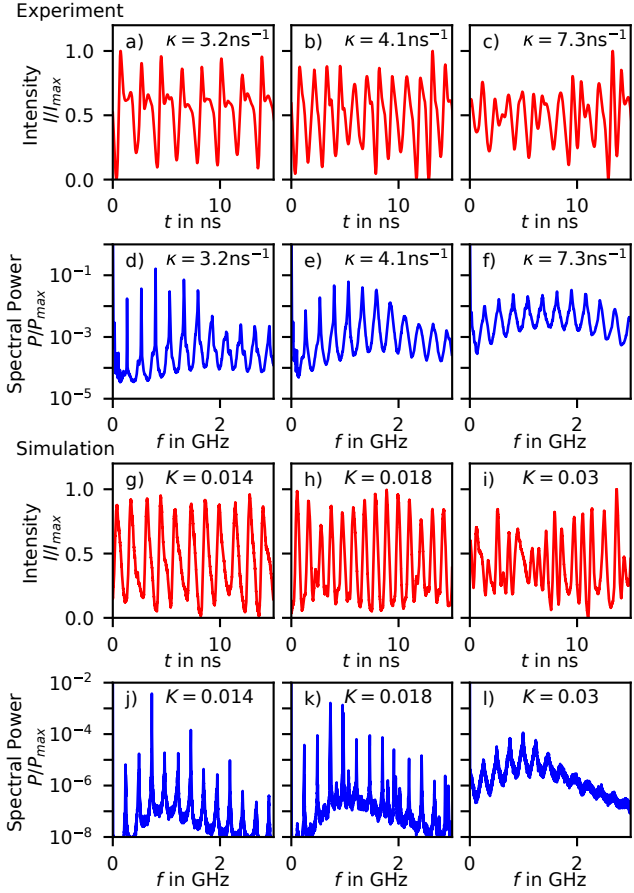}
		\caption{a)-f) Experimentally measured time series and corresponding normalized RF-spectra at a pump current of $J/J_{th} = 1.4, 1.4, 1.25$ from left to right. g)-l) Simulated data that qualitatively reproduces the experimentally obtained time series and spectra. With pump currents of $J/J_{th} = 1.09, 1.11, , 1.11$ from left to right. All other Parameters as given in Table \ref{Tab1}.} 
		\label{Fig3}
	\end{figure}

	When introducing single-mode feedback to the SMLQD laser device we find that above a critical feedback strength the CW dynamics switches to a periodic orbit characterized by a pulsed-like laser output as shown in real time oscilloscope traces of the laser intensity $I = |E|^2$ shown in Fig.\ref{Fig3}a)-b). The measurements were taken at increasing feedback rates of $\kappa = 3.2$\,ns$^{-1}$ and $\kappa = 4.1$\,ns$^{-1}$, where $\kappa$ is the effective feedback rate calculated from the differences in the laser P-I curve of free-running and feedbacked laser device (see appendix \ref{AppK}). It is possible to identify a transition from a pulsed-like regular periodic orbit to a more irregular/chaotic behaviour at increasing feedback strength. This is also supported by the corresponding radio-frequency (RF) spectra shown in Fig.\ref{Fig3}d)-f). At low frequencies ($f < 1.5$\,GHz) we clearly obtain mode-peaks several magnitudes above the background. These modes are separated by $\approx 1/\tau = 0.26$\,GHz, i.e. the inverse of the feedback time ($\tau =3.8$\,ns). At higher feedback strengths the peak power of the modes decreases drastically relative to the background (see Fig.\ref{Fig3}e). This evolution continues when increasing the feedback-strength even further as shown in the RF spectrum in Fig.\ref{Fig3}f) measured at $\kappa = 7.3$\,ns$^{-1}$, which corresponds to the irregular/chaotic time-series in Fig.\ref{Fig3}c). 
	The transition of the dynamics with increasing feedback strength can qualitatively be reproduced utilizing the introduced SMLQD laser model. At low feedback strength we find regular pulsed-like solutions, which are characterized by strong locking peaks at the dominant frequencies in the RF-spectrum as indicated in Fig.\ref{Fig3}g)j). With increasing feedback strength the locking is less pronounced (see Fig.\ref{Fig3}l)) and we therefore find chaotic/irregular solutions as shown by the time-series in Fig.\ref{Fig3} i). \\
	This phase-locking effect, in which the phases of ECMs and the relaxation oscillation lock, was analogously found investigating quantum dot and quantum well lasers \cite{TYK16,KEL17b} as well as opto-electronic feedback \cite{KOV19,ISL20}. Our findings also support the claim that a high damping, also typical for SMLQD lasers \cite{LIN16}, is essential for this locking effect to occur at intermediate feedback strengths without generating chaotic dynamics \cite{ABD03,KEL17b}. Furthermore, experimental and theoretical findings as well as this work suggest an intermediate feedback length of $1$\,ns $< \tau <$ 10\,ns close to the relaxation oscillation period being beneficial this type of locking effects \cite{PIE01,TYK16,KEL17b,KIM15,TAB04}.
	\begin{figure}
		\includegraphics{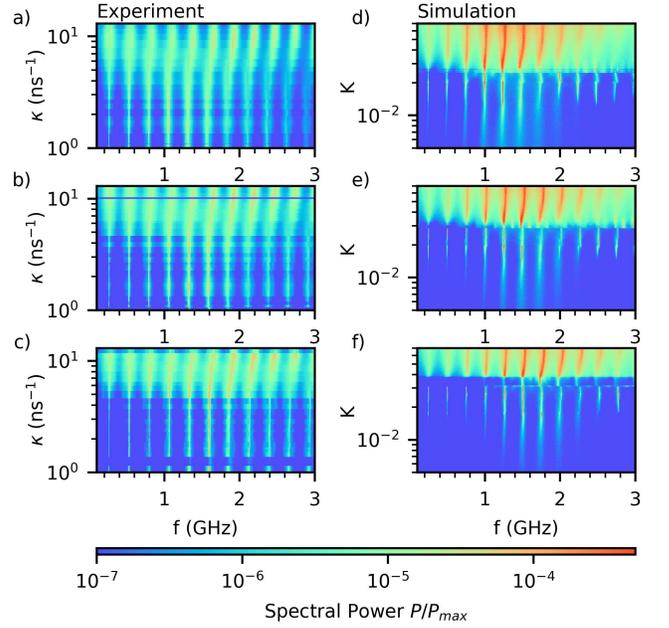}
		\caption{a)-c) Experimentally obtained RF-spectra at different feedback rates $\kappa$ (see Appendix \ref{AppK}). The normalized spectral power is given by the color-code. The pump current was increased from top to bottom according to: $J/J_{th} = 1.33, 1.41, 1.48$, d)-f)  RF-spectra generated by direct numerical integration of the SMLQD model at different feedback strengths $K,$ with increasing pump currents from top to bottom $J/J_{th} = 1.19, 1.26, 1.32$. The spectral power of the experimental data was multiplied by $10^3$ for a better comparability. All other Parameters as given in Table \ref{Tab1}.}
		\label{Fig4}
	\end{figure}
	For a more in depth analysis of the underlying dynamical transitions we continuously decrease the feedback rate starting from the maximum achievable value and measure the RF-spectra of the laser output at each feed-back strength. The result of the experimentally obtained data for different pump currents is indicated in the 2D plots in Fig.\ref{Fig4}a)-c), with the color-code indicating the spectral power. One can note narrow locking peaks several magnitudes above the background in an intermediate region of feedback strengths between $\kappa = 1.5$\,ns$^{-1}$ and $\kappa = 4$\,ns$^{-1}$, which indicates the presence of regular periodic orbits in this regime. At higher feedback-strengths, the background increases and the peaks broaden, suggesting chaotic/irregular dynamics. With increasing pump currents from a) to c), the onset of the chaotic dynamics shifts to higher feedback strengths. Applying the theoretical model, we can also predict this feedback effect, as the calculated spectra show the same behaviour (Fig.\ref{Fig4}d)-f)). At low feedback ($ 0.01 < K < 0.05$) sharp and well separated peaks of the oscillatory modes can be found. However, when increasing the feedback strengths, the modes smear out and chaotic behaviour emerges, with the boundary between periodic orbits and chaotic dynamics slightly shifting to higher $K$ with increasing pump current. 
	\begin{figure}
		\includegraphics{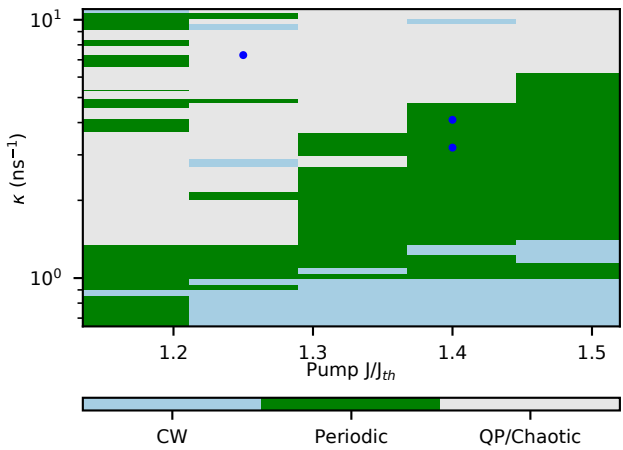}
		\caption{Experimentally obtained map of the laser dynamics in the ($J,\kappa$)-plane. The dynamical regimes are distinguished according to the RF spectra: continuous wave dynamics (CW) are characterized by a low radio frequency signal, periodic dynamics are characterized clear locking peaks with a low background (Fig.\ref{Fig3}d), whereas chaotic/quasi periodic (QP) dynamics refer to broadened spectral peaks and a high background (Fig.\ref{Fig3}f). The blue dots correspond to the parameters used for the time-series in Fig.\ref{Fig3}. }
		\label{Fig9}
	\end{figure}
	\ \\ By performing further sweeps of the feedback-rate whilst measuring the RF-spectra for different pump currents we can also get a broader overview of the dynamics in the 2D ($J,\kappa$)-plane. We distinguish the different dynamical regimes according to their RF-spectra and cross-check with the measured time-series of the laser output. The result is presented in Fig.\ref{Fig9}. We find that for all pump currents the device turns from CW to a periodic output state. As described before, the laser destabilizes into a chaotic state at higher feed-back rates (grey areas in Fig.\ref{Fig9}). The transition to chaos either mediated by a period doubling cascade or quasi periodic transition is a well known feedback effect \cite{MOR90,ABD03,KIM15}, whereas the locking behaviour at intermediate feed-back strengths is the same effect found for quantum dot and quantum well lasers under optical and electro-optical feedback \cite{TYK16,KEL17b,ISL20}. 
	\newpage
	\section{Bifurcation Analysis}
	In order to understand the generation mechanisms of the different dynamical regimes, we investigate the role of the gain $g$ the pump current $J$ and the $\alpha$-factor by performing direct numerical integrations as well as path continuation utilizing DDE-biftool \cite{ENG02}. 
	\begin{figure}[b]
		\includegraphics{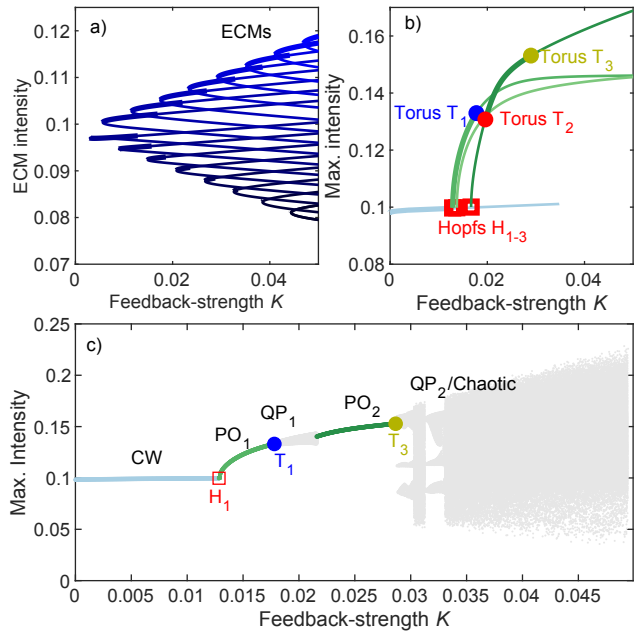}
		\caption{a)b) 1D-Bifurcation diagram showing the maximum intensity as a function of the feedback-strength $K$ for the continuous wave ECM solutions in a) as well as the CW solution (cyan) and periodic orbits (green) b). Thick (thin) lines denote stable (unstable) dynamics. The three periodic orbits in b) are born in subsequent Hopf bifurcations (H$_{1-3}$) along the CW branch and can be distinguished according to the oscillation period of the born solutions, separated by integer multiples of the inverse feedback time $n/\tau = 0.26$\,GHz.  The Tours bifurcations $T_{1-3}$ are marked by the coloured dots. Panel c) indicates the dynamics obtained performing a direct numerical integration (up-sweep), showing the unique intensity maxima found in 200 external cavity round-trips at each $K$. Chaotic/quasi-periodic regimes are indicated in grey. All parameters as in Table \ref{Tab1} and $J/J_{th} = 1.19$. } 
		\label{Fig5}
	\end{figure}
	\ \\ We find that at the investigated feedback-strengths already a high number of external cavity modes (ECMs) is stable as indicated by the 1D bifurcation diagrams displaying the intensity of the ECMs as a function of the feedback-strength in Fig.\ref{Fig5}a). This behaviour is expected from various analyses of the Lang-Kobayashi style feedback implementation \cite{LAN80b,LYT97,ROT05}. The stability of several ECMs at these feedback strength suggests a high influence of the feedback on the laser dynamics. Hence, when continuing the CW solution of the solitary laser, we find that after several ECMs have become stable ($K > 0.015$) subsequent Hopf bifurcations appear leading to a pulsed-like laser output (see green branches in Fig.\ref{Fig5}b). Compared to the case of standard laser diodes \cite{PIE01,TAB04}, the Hopf bifurcations appear very close after another on the CW branch and the stable periodic orbits are not connected to the ECMs or born from bifurcations along them. Therefore this behaviour is similar to the case of opto-electronic feedback \cite{ISL20}. However, periodic orbits born from the Hopf bifurcations along the ECMs also exist. These turn unstable very close to the bifurcation point in the investigated parameter regime. \\
	The frequency of the periodic orbits born in the Hopf bifurcations along the CW branch is separated by the inverse of the feedback time $1/\tau = 0.26$\,GHz. However, these bifurcations do not necessarily appear ordered with respect to the frequency as an increase in pump current leads to a reordering of the Hopf bifurcations along the CW branch, which is discussed in the subsequent section. We also investigate the direct numeric integration of the system sweeping up the feedback-strength, i.e. increasing the feedback-strength in small steps and using the dynamics of the previous step as the initial condition of the next to emulate the experiment. The result is shown in Fig.\ref{Fig5}c). In between the periodic orbits with different frequency, a quasi-periodic/chaotic regime is stable, resulting from the T$_1$ Torus bifurcation of the first periodic orbit. However, this regime destabilizes at slightly higher feedback-strength ($K \approx 0.022$) and the system stabilizes on the second regular periodic orbit, as expected from the path continuation in Fig.\ref{Fig5}b). As for laser diodes with 3D gain material, a route to chaos can be seen at higher feedback strength \cite{MOR90b,MOR92}.
	\ \\ In order to reproduce the experimental results shown in Fig.\ref{Fig9} and get a deeper insight into the dynamics, we also perform numerical scans of the dynamics in the ($J,K$) plane. The results are presented in Fig.\ref{Fig6}, where different dynamics can be distinguished according to the color-code as before. Generally, we can reproduce the influence of the feedback. At the investigated pump currents, the laser firstly transitions from the CW state to a periodic regime and then turns to a chaotic output at higher feedback-strengths. Moreover, we find that the general bifurcation scenario related to a variation in feedback-strength is not drastically changed when altering the pump current. The pump current for which the 1D bifurcation diagram is shown in Fig.\ref{Fig5}b), is indicated by the vertical black dotted line in Fig.\ref{Fig6}a). However, the first periodic orbit (green areas in Fig.\ref{Fig6}a) stabilizing from the CW solution (light blue) changes as the Hopf bifurcations along the CW branch swap their position (H$_1$, red lines in Fig.\ref{Fig6}a)), when the pump current is altered. Hence, periodic orbits with different frequencies stabilize at different pump values, which is indicated by the varied shading of the periodic orbit regimes (green). As mentioned before, the frequencies of the POs differ by integer multiples of the inverse feedback time. We note that the periodic orbits stabilizing first at lower pump currents, stabilize second at higher pump currents as indicated by the out-stretching green areas at high feedback-strength. The shift of the transition point between CW lasing and pulsed/chaotic behaviour to higher feedback-strengths with increasing pump was also found for other Lang-Kobayashi type feedback systems \cite{ABD03}. \\
	The Torus bifurcations which destabilize the first periodic regime when increasing the feedback strength $K$ (see T$_1$ Torus Fig.\ref{Fig5}b) are indicated as dashed lines in Fig.\ref{Fig6}b). They are born at the intersection points two Hopf lines (see circles in Fig.\ref{Fig6}a). One of the Hopf bifurcations is giving birth to the periodic orbit destabilized by the T$_1$ Torus bifurcation. The other Hopf line is related to the Hopf bifurcation point existing closest to the lasing threshold and has the fundamental ECM frequency. The T$_1$ bifurcations in turn give birth to quasi periodic solutions (grey areas) in between the regular periodic orbits ($0.125 < K < 0.038$). At higher feedback-strength these quasi periodic regimes become unstable and a second regular periodic orbit stabilizes. The stability boundary of these periodic orbits is given by two torus bifurcations as indicated by T$_2$ and T$_3$ in Fig.\ref{Fig5}b). The bifurcation lines of these Torus bifurcations are shown as red and yellow dashed lines in Fig.\ref{Fig6}c,d).
	In order to numerically unravel the lower stability boundary (T$_2$) a down-sweep was computed in Fig.\ref{Fig6}d). The quasi periodic regime stabilizes at a lower feedback strength than the second periodic orbit and therefore becomes stable first in the upsweeps shown in Fig.\ref{Fig6}a-c). 
	\section{Period Change}
	\begin{figure}
		\includegraphics{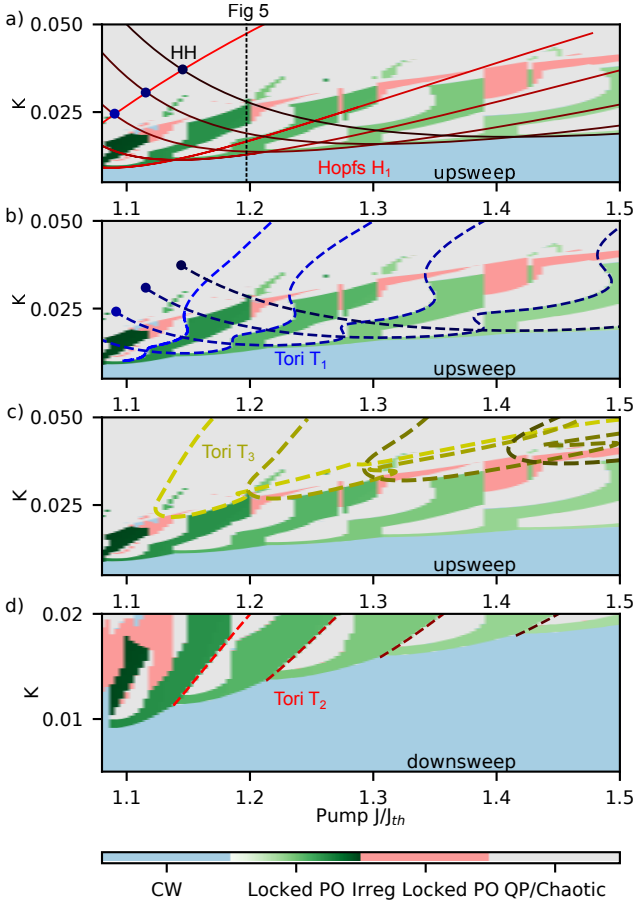}
		\caption{2D bifurcation diagrams in the ($J,K$) parameter plane, with the different dynamics distinguished by the color-code. The green shading of the regular periodic solutions (pulsed) refers to the different oscillation frequencies. The red areas describe pulsed dynamics which is slightly irregular similar to the time series shown in Fig.\ref{Fig3} h). Panel a)-c) show an upsweep in the feedback strength, whereas a downsweep is displayed in d). The red lines in a) are the Hopf bifurcation lines corresponding to the first Hopf $H_1$ bifurcation along the CW branch (see Fig. 5) found at different pump currents. The crossing points (HH - blue circles top left) of these Hopf lines  give birth to the torus lines (dashed lines) shown in b), representing the upper stability boundary $T_1$ of the first periodic orbit PO$_1$, as shown exemplary for $J/J_{th} = 1.19$ in Fig.\ref{Fig5}b). The dashed lines in c) and d) represent the upper and lower stability boundary of the second periodic solution PO$_2$, i.e. the $T_2$ and $T_3$ Torus in Fig.\ref{Fig5}b). All parameters as in Table \ref{Tab1}.} 
		\label{Fig6}
	\end{figure}
\begin{figure}
		\includegraphics{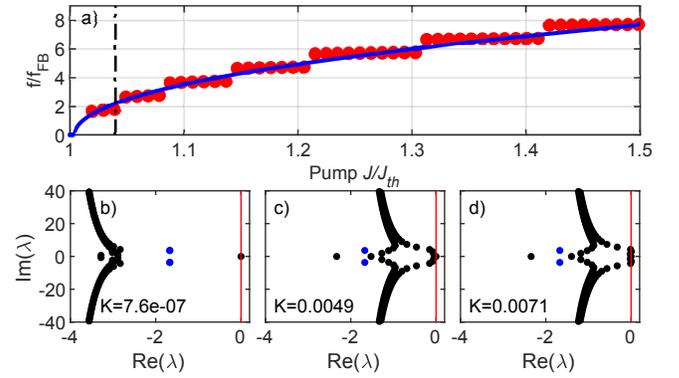}
		\caption{a) Relaxation oscillation frequency (blue line) and frequency of the periodic orbit (red dots) generated by the first Hopf bifurcation ($H_1$) along the CW branch (cf. Fig.\ref{Fig5}) as a function of the pump current and normalized to the feedback frequency $f_{FB} = 1/\tau$. b)-d) Real and imaginary parts of the Lyapunov exponents along the CW solution for increasing feedback strengths, with d) showing the Hopf point. The Eigenvalues corresponding to the relaxation oscillation frequency are marked in blue.  The was pump current is set to $J/J_{th} = 1.04$ as marked by the vertical black line in a). All parameters as in Table \ref{Tab1}. } 
		\label{Fig7}
	\end{figure}
\begin{figure}
		\includegraphics{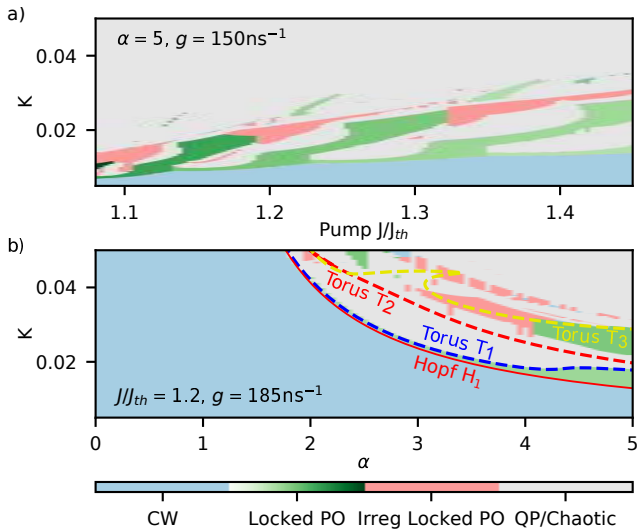}
		\caption{a) 2D bifurcation diagrams in the ($J,K$) parameter plane, with the different dynamics distinguished by the color code as in Fig.\ref{Fig6}, but for a lower gain $g = 150$. b) 2D bifurcation diagrams in the ($\alpha,K$) parameter plane. The solid line marks bifurcation line of the first Hopf $H_1$, the dashed lines correspond to the Torus bifurcations $T_{1-3}$ as displayed in Fig.\ref{Fig5}b) (1D cut at $\alpha = 5$). The pump current is $J/J_{th} = 1.19$, all other parameters as in Table \ref{Tab1}. } 
		\label{Fig8}
	\end{figure}

	In the following, we investigate the frequency of the first stable periodic orbit and its relation to the relaxation oscillation frequency (RO). We therefore find the frequency of the periodic orbit at the first Hopf point in $K$ at increasing pump currents and also find the relaxation oscillation frequency for the respective pump. The periodic orbit's frequency is indicated by red dots in Fig.\ref{Fig7}a), whereas the dependence of the ROs is given by the blue line. We obtain the results by extracting the Eigenvalue spectrum of the CW steady state applying a linear stability analysis using DDE biftool. The imaginary part of the non-trivial eigenvalues $\lambda$ existing at $K=0$ corresponds to the relaxation oscillation frequency via the relation $f_{RO} = \Im(\lambda)/2\pi$, whereas the imaginary part of the eigenvalue pair crossing the imaginary axis at $K_{Hopf}$ relates to the Hopf frequency. The Eigenvalue spectrum is displayed in in Fig.\ref{Fig7}b-d), for three feedback-strengths at the pump current indicated by the black line in Fig.\ref{Fig7}a). The eigenvalue pair corresponding to the RO is marked in blue. \\
	The RO shows a square root dependence (see Fig.\ref{Fig7}a)) on the pump current, whereas the change of the stabilizing PO frequency omits a staircase behaviour in which the frequency always increases in steps of the feedback/external cavity frequency $f_{FB}$. Hence, we can deduce that the  Eigen value pair with the frequency closest to the RO leads to the stabilization of the periodic orbit via a Hopf bifurcation. Due to the coupling between the bulk reservoir and the SMLQD states the investigated device inhibits a high damping, which was argued to be a key property necessary for the locked periodic orbits to occur \cite{KEL17b}. Hence, the presented explanation of the step-wise frequency change also applies to the staircase scenario found in the measurements obtained using a quantum dot device in \cite{KEL17b}. 
	One can also understand the discussed locking effect under the aspect of a destabilization of the CW lasing. Above a critical feedback-strengths the excess depletion of the gain resulting from the feedbacked intensity cannot be compensated and therefore the CW state destabilizes to a periodic orbit. With increasing pump current the relaxation oscillation frequency shifts to higher values and therefore higher periods are favoured as shown before. As the RO is strongly influenced by the gain factor $g$, the latter also influences the locking frequency. This is supported by the 2D bifurcation diagram in the ($J,K$)-plane in Fig.\ref{Fig8}a) calculated with a reduced $g$. Here the areas of single periodic orbits stretch over a larger relative pump current and the transitions to higher oscillation frequencies are shifted to higher pump currents. The reason for that lies within a slower increase of the RO frequency with increasing pump current, due to the reduced gain. \\
	The investigation of laser diodes with quantum dot, quantum well or bulk gain all show that a high $\alpha$ factor is very beneficial for the locking oscillations to occur \cite{TYK16,KEL17b,PIE01}. We obtain similar findings as the bifurcation diagram in the $(\alpha,K)$-plane (Fig.\ref{Fig8}b) shows that the periodic orbits are completely lost for feed-back levels $K < 0.05$ if the $\alpha$ factor is reduced below $\alpha < 2$. This results from the Hopf bifurcation H$_1$  and the torus bifurcation T$_1$, enclosing the stable region of periodic locked orbits, strongly shifting to high feedback-strengths with decreasing $\alpha$-factor. Furthermore, these bifurcations approach each other as $\alpha$ is scaled down and hence reduce the region of stable periodic orbits. At this point we highlight that a high $\alpha$ is required by the theoretical model to correctly model the properties of the SMLQD gain system. Additionally, we note that the second regime of regular periodic pulsed-like dynamics turns to an irregular behaviour at decreasing $\alpha$ values as indicated in Fig.\ref{Fig8}b). 
	\section{Conclusion}
	To conclude, we have presented the experimental and theoretical investigation of a semiconductor laser with strong amplitude-phase coupling subject to optical feedback at an intermediate feedback length of $\tau = 3.8$\,ns, using submonolayer quantum dots as the active medium. We present a new theoretical model for a SMLQDs laser diode subject to optical feedback, which correctly describes the damping and relaxation oscillations by the coupling of the bulk reservoir and SMLQD states. We validate the essential features of the relevant time-scales by comparison to previous \cite{HER16,LIN16} and the experimental results. Utilizing this model we can reproduce an experimentally obtained phase-locking effect at intermediate feed-back strengths which leads to a periodic pulse like laser output. We complement our experimental findings by a theory that allowed us to explore the step-wise frequency change of the periodic orbits with increasing pump current. Specifically, our results show that the step-size always is related to the inverse feedback-time and is induced by the eigenvalue pair closest to the relaxation oscillation frequency stabilizing the periodic orbit. This can be generalized to previous experiments on quantum dot and quantum well lasers \cite{TYK16,KEL17b}, as the high damping of the devices seems to be the most important requirement for the locking effect, which is predominantly induced by the feedback. Comparing our findings to existing literature also suggests that a high $\alpha$-factor and an intermediate feedback lengths $1$\,ns~$< \tau < 10$\,ns close to the relaxation oscillation frequency seem to be beneficial for these type of oscillations to occur. Furthermore, we perform an in-depth bifurcation analysis showing that Torus bifurcations lead to a stabilization of further periodic branches at higher feed-back strengths and the intermediate transition to a quasi-periodic laser output. The generalization of the underlying feedback effect with a step-wise change in period can be of interest for the future development of tunable photonic micro-wave sources.
	
	
	The authors thank the Deutsche Forschungsgemeinschaft (DFG) within the frame of the SFB787 (all) and SFB910 (J.H., K. L.) for funding.
	\appendix
	\section{Modelling the equilibrium distribution}
	\label{AppRho}
	The SMLQD equilibrium occupation probability $\rho_{eq}$ is determined by a quasi-Fermi distribution:
	\begin{align}
	\rho_{\mathrm{eq}}(t) = \left[ \exp \left( \frac{\bar{\varepsilon}_{\mathrm{SML}} - E_{\mathrm{F}}(t)}{ k_{\mathrm{B}} T} \right)  + 1 \right]^{-1},
	\end{align}
	where $\bar{\varepsilon}_{\mathrm{SML}}$ is the mean SMLQD confinement energy with respect to the bulk reservoir band edge, $m^{*}$ is the effective mass, $T$ is the lattice temperature, is $E_{\mathrm{F}}$ is the quasi-equilibrium Fermi level, which is determined via the Pade approximation \cite{COL95}:
	\begin{align}
	E_{\mathrm{F}} =& k_{\mathrm{B}} T \{  \ln[r(t)] + A_1 r(t) + \nonumber \\
	&\left[ K_1 \ln[1 + K_2 r(t)] - K_1 K_2 r(t) \right] \}.
	\end{align}
	with $r = n(t)/n_{\mathrm{C}}$ where the effective density of states  $n_{\mathrm{C}}$ is given by 
	\begin{align}
	n_{\mathrm{C}} = 2 \left( \frac{m^* k_{\mathrm{B}} T}{2 \pi \hbar^2} \right)^{3/2}.
	\end{align}
	The coefficients are $A_1 = 1/\sqrt{8}$, $A_2 = -4.95009 \times 10^{-3}$, $K_1 = 4.7$ and $K_2 = \sqrt{2 |A_2|/K_1}$.
	\section{Maximum Feedback rate}
	\label{AppK}
	In order to give a measure of the effective feedback rate $\kappa$, we calculate the maximum effective feedback rate $\kappa_{max}$ for the investigated etalon mode as follows:
	\begin{align}
	\kappa_{max} = \frac{(1-R_{GaAs})}{\tau_{LD}}\frac{E_{ph}}{e}\bigg[\Big(\frac{\delta I}{\delta P}\Big)_{EC} - \Big(\frac{\delta I}{\delta P}\Big)_{LD}  \ \bigg],
	\end{align}
	where $R_{GaAs} = 0.33$ is the reflectively of the GaAs facette, $E_{ph} = 1.16$\,eV is the photon energy, $\tau_{LD} = 12$\,ps is the round-trip time of the laser diode,  $e$ is the the electron charge, and $\Big(\frac{\delta I}{\delta P}\Big)_{LD,EC}$ refer to the fitted linear slope of the P-I laser curve of the free-running laser doide ($LD$) and the laser under maximum feedback ($EC$). By utilizing the set attenuation of the neutral density filter $\beta$ in dB, we then determine the effective feedback-rate as follows: 
	\begin{align}
	\kappa = \kappa_{max} 10^{\beta /20}.
	\end{align}

	\clearpage
	\bibliography{export}
	
	\bibliographystyle{unsrt}
	\clearpage
	\appendix

\end{document}